\documentstyle[aps]{revtex}
\begin{document}
\draft
\title{Parity-violating Kalb-Ramond-Maxwell interactions and CMB anisotropy in 
a braneworld}
\author{Debaprasad Maity\footnote{E-mail: tpdm@iacs.res.in}$^a$, 
Parthasarathi Majumdar\footnote{On deputation from the Institute of 
Mathematical Sciences, Chennai 600 113, India; E-mail: 
partha@theory.saha.ernet.in}$^b$ and Soumitra
SenGupta\footnote{E-mail: tpssg@iacs.ernet.in}$^a$}
\address{$^a$ Department of Theoretical Physics, Indian Association 
for the Cultivation of Science, Kolkata 700 032, India}
\address{$^b$ Theory Group, Saha Institute of Nuclear Physics, AF/1 
Bidhannagar, Kolkata 700 064, India}
\maketitle
\begin{abstract}

Following up on a recent paper by two of us (DM and SS),
demonstrating the large enhancement in observable optical activity in
radiation from high redshift sources arising from the string-based
coupling of bulk Kalb-Ramond field to the Maxwell Chern Simons
three-form on the brane in a Randall-Sundrum braneworld, we exhibit
here a similar enhancement in parity-violating
temperature-polarization correlations, yet unseen, in the CMB
anisotropy due to a generalized parity-violating Kalb-Ramond axion-photon
interaction proposed earlier by one of us (PM). The non-observation
of such correlations in CMB anisotropies would necessitate unnatural
fine tuning of the Kalb-Ramond axion parameters. As a stringy realization of 
Randall-Sundrum braneworld scenario is yet 
to be understood properly , our work indicates the need of
a careful investigation to establish the connection between string-based 
phenomenological models and the Randall-Sundrum braneworld scenario.
\end{abstract}
\vglue .3in

The ubiquity of the Kalb-Ramond (KR) antisymmetric tensor field in massless spectra
of all closed string theories makes it, along with the dilaton, a special low energy
degree of freedom naturally occurring in a stringy world. On the other hand, the
standard strong-electroweak theory does not appear at all to need such a field for
an adequate description of TeV scale phenomena. Therefore any observational evidence
at low energies of KR fields would point to physics beyond the standard
strong-electroweak theory. It stands to reason that such an evidence would be
indicative of an underlying fundamental theory whose tree level spectrum must
resemble closely that of some sort of the string (M-) theory. The focus of this 
paper is precisely on such possible indirect observational evidence of KR fields. 

The KR field is known in heterotic string theory \cite{gsw} to have a
coupling, induced at the quantum level, to the Yang-Mills
Chern-Simons three-form. Basically, it entails augmenting the KR
three-form field strength $H \rightarrow H + \Omega_Y$ where
$\Omega_Y$ is the Yang Mills Chern Simons three-form. This coupling
originates from the requirement of quantum consistency (cancellation
of anomalies) and $N=1$ supersymmetry of the theory. The interaction
survives compactification to four dimensions, leading to the low
energy effective interaction \cite{ms} of the form $S_{int} \sim
(1/M_P) \int d^4x \phi_H F \cdot ^*F$, where $\phi_H$ is the
pseudoscalar field (KR axion) dual to the KR field strength, $F$ is the
Maxwell field strength and $M_P$ is the Planck mass. Assuming a
homogeneous background KR field with little backreaction on
electromagnetic fields, this coupling has been shown
\cite{kmss},\cite{kmss2} to lead to an optical activity in
electromagnetic radiation from high redshift sources in the form of
rotation of the plane of polarization by an angle $\Delta \Phi= h
t(z)$. Here $h$ is essentially $ d\phi_H/dt$, and $t(z)$ is the
`lookback' time as a function of the redshift $z$. Observational
upper limits on $\Delta \Phi$ \cite{ward} bound $h$ from above by a
number of $O(10^{-5})$ in units of Planck mass.

A different possibility seems to appear if the same interaction is considered in a
higher dimensional low energy effective model proposed about five years ago - the
Randall-Sundrum (RS) scenario \cite{rs}. Although so far this model has not been
derived in detail from string (M-) theory, compactifications with non-vanishing
fluxes of $p$-form gauge fields in the Ramond-Ramond sector appear to generate
scenarios similar to the RS model \cite{gid,kleb}. Furthermore, there is as 
yet no
indication that the RS model contradicts any basic tenet of M-theory
compactifications. 

In the RS scenario, massless closed string modes like the KR field, (alongwith
gravitons) propagate in the bulk (i.e., in all spacetime dimensions including the
compact one) while all the standard model fields (open string modes) are confined to
the brane. Of course, there are interactions between closed and open string modes,
as required by string consistency. In this paper we shall concentrate on one such
particular interaction, namely the interaction between the KR field (in the bulk) 
and the Maxwell field on the visible brane. In this context, the heterotic string 
has a special significance since its spectrum in principle accommodates all fields 
of the standard model, including the Maxwell field (after appropriate 
compactification). KR-Maxwell couplings occurring in the heterotic string are 
therefore of particular relevance here. As already mentioned above, the KR field 
strength tensor couples to the Chern Simons three form of the Maxwell field in 
the heterotic string. 

Recently, two of the authors \cite{mas} have considered such an interaction of the
Maxwell Chern Simons three-form to a homogeneous KR background within the RS
braneworld scenario.  The analysis shows that the coupling is enhanced by the {\it
inverse} of the warp factor in the five dimensional RS metric. We may recall that
this warp factor accounts for the exponential suppression which ensures naturalness
of the electroweak theory by making the Higgs mass stay close to the weak scale; it
is a number of $O(10^{-16})$.  With the inverse of this factor appearing in the
KR-Maxwell coupling, the optical rotation of radiation from large redshift sources
increases enormously. To reconcile the resultant theoretical angle of rotation with
the (non-)observations, one needs to fine tune the parameter $h$ (in units of the
Planck scale) to a precision of less than one part in 10$^{15}$ - certainly {\it
unnatural} from the standard `naturalness' perspective.

In this paper, we extend the analysis of \cite{mas}, so as to be
applicable to the anisotropy of the Cosmic Microwave Background (CMB)
through the introduction of spatial parity (P-) violation. It has
been noted several years ago \cite{liu} that certain non-vanishing
multipole moment correlations between the temperature anisotropy and
polarization of the CMB could appear, if in the interaction described
above, the KR axion was to be replaced by a homogeneous {\it scalar}
field, leading to P-violation. Such an interaction appears in a
proposal in \cite{pm} in the form of an extended augmentation, of the
odd parity KR field strength, by a term which has even parity. As we
shall explain below, this can produce P-violating multiple moment
correlations of temperature anisotropy and polarization of the CMB,
without having to change parity of the free KR field. With this
additional augmentation of the KR field strength, the analysis of
\cite{mas} now carries over easily to the CMB case, as we proceed to
describe.

The angular distribution of the temperature anisotropy of the CMB has the
well-known expansion in spherical harmonics \cite{cowh}
\begin{eqnarray}
{\Delta T \over T}({\mathbf 
n})~=~\sum_{l,m}a^T_{lm}~Y^T_{lm}({\mathbf n})~.
\label{tani}
\end{eqnarray}
The polarization of the CMB is expressed in terms of a 2 $X$ 2
traceless symmetric tensor ${\cal P}_{ab}({\mathbf n})$ whose
components are the Stokes parameters. This tensor can be decomposed
into its irreducible `gradient' (or $E$) and `curl' (or $B$) parts
which have opposite spatial parity. The angular distribution of this
polarization tensor can thus be expressed in terms of matrix
spherical harmonics as \cite{liu}

\begin{eqnarray}
{\cal P}_{ab}^E({\mathbf n})&~=~&\sum a_{lm}^E~Y_{lm,ab}^E({\mathbf 
n})~ \nonumber
\\
{\cal P}_{ab}^B({\mathbf n})&~=~&\sum a_{lm}^E~Y_{lm,ab}^B({\mathbf n})~.
\label{pol}
\end{eqnarray}

One defines correlations of the multipole moment coefficients
$a_{lm}^X~,~X = T,E,B$ as 
\begin{eqnarray}
C_{l}^{XX'} \equiv \langle a_{lm}^X~a_{lm}^{X'} \rangle ~. 
\label{cor} 
\end{eqnarray}
Clearly, correlations such as
$C_l^{XX}$ all preserve P, as do correlations like $C_l^{TE}$, while
correlations like $C_l^{TB}$ or $C_l^{EB}$ are obviously P-violating,
requiring an explicitly P-violating interaction as mentioned earlier.
The optical activity described earlier implies that if a correlation
like $C_l^{TE}$ does indeed arise due to reionization or otherwise,
the passage of the Thompson scattered photons through the KR
background will produce the P-violating correlations $C_l^{TB}$
through the rotation \cite{liu}
\begin{eqnarray}
C_l^{TB} = C_l^{TE}~sin ~2 \Delta \Phi~. \label{pvcor}
\end{eqnarray}
Such a parity violating effect has been shown to arise from a 
phenomenological action of the type \cite{liu}
\begin{eqnarray}
S_{int} ~\sim ~\int~\phi~ F \cdot ^*F~ , \label{pvkam}
\end{eqnarray}
where $\phi$ is an even parity scalar. 

Let us now see how such P-violation may possibly arise within a
string theoretic framework. The standard augmentation of the KR field
strength is given, at the effective 4d field theory stage, by
\begin{eqnarray}
H_{\mu \nu \lambda} \rightarrow {\tilde H}_{\mu \nu \lambda} ~ \equiv~  
H_{\mu \nu \lambda} ~+~ {1 \over 3M_P} A_{[\mu} F_{\nu \lambda]}
~.\label{htil}
\end{eqnarray}
We now propose, following \cite{pm}, the extended version of
augmentation of the KR field strength
\begin{equation}
{\tilde H}_{\mu \nu \lambda} ~=~ H_{\mu \nu \lambda}~+~{1 \over 
3M_P}~\left[
\zeta_+~A_{[\mu}~F_{\nu \lambda]}~+~\zeta_-~A_{[\mu} ~^*F_{\nu 
\lambda]} \right]~,
\label{htild}
\end{equation}
where, $\zeta_{\pm}$ are $O(1)$ dimensionless real numbers. The
first term is the standard Chern Simons augmentation, while the
second is our extension; its main purpose is to contaminate the odd
parity KR field strength with a small (due to Planck mass
suppression) even parity impurity. Since ${\tilde H}^2$ appears in
the effective action, it must be invariant under both the KR and
Maxwell gauge transformations. The latter invariance requires that
the unaugmented $H_{\mu \nu \lambda}$ be assigned a Maxwell $U(1)$
gauge transformation
\begin{eqnarray}
\delta_U H_{\mu \nu \lambda}~=~- {1 \over 3M_P} \partial_{[\mu}
\omega~\left [\zeta_+ F_{\nu \lambda]}~+~\zeta_- ^* F_{\nu 
\lambda]} \right] ~. 
\label{gaug}
\end{eqnarray}

An immediate consequence of this gauge transformation is that the
$H_{\mu \nu \lambda}$ now can no longer be thought of as a parity
eigenstate, and therefore neither can its dual spinless field
$\phi_H$. In other words, one can decompose 
\begin{eqnarray}
\phi_H = \phi_H^{(+)}~ +~ \phi_H^{(-)} \label{dec}
\end{eqnarray}
where $\phi_H^{(\pm)}$ has parity $\pm$. It is now easy
to show \cite{pm} that the interaction part of the effective action
contains a P-violating term 
\begin{eqnarray}
S_{int}^{\mathrm PV} \sim M_P^{-1} \int \phi_H^{(+)}~F \cdot ^*F~, 
\label{pvio} 
\end{eqnarray}
i.e., precisely the sort of coupling that leads to the P-violating 
correlations $C_l^{TB}$ in the CMB anisotropy. We note in passing 
that the P-violating interaction also violates time reversal 
invariance (T) so as to leave PT invariant \cite{liu}. 

A few remarks about the extended augmentation are in order: first of
all, unlike the Chern Simons augmentation, it is not topological
(nonlocal) in nature and hence does not introduce any new anomalies.
It is most likely a local counterterm at the quantum level. Such
counterterms commonly occur in local field theories so long as
they obey all gauge invariances.  Their determination requires inputs
(`renormalization conditions') related to experiment. 

Second, it can easily be embedded in a supersymmetric theory at the 
$N=1,D=4$ level. To see this, recall that the locally 
supersymmetric Maxwell action in $N=1,D=4$ superspace has the form
\begin{eqnarray}
S_{Max} \sim \int d^4x~d^2 \theta~\chi({\cal 
S})~W^{\alpha}~W_{\alpha}~+~\rm{h.c.}
~, \label{susy}
\end{eqnarray}
where the gauge kinetic function $\chi({\cal S})$ is an analytic
function of the chiral superfield ${\cal S}$ whose top component is
the complex scalar $\phi_D + i \phi_H$. If this gauge kinetic
function $\chi({\cal S}) \rightarrow \zeta_+ \chi({\cal S}) +
\zeta_- {\tilde \chi}(i{\cal S})$, the component action indeed has
the interaction terms discussed above. With such features one expects 
the extension proposed above to be consistent with quantum string 
theory. 

As argued in \cite{pm}, the calculation of the angle of rotation of
the polarization plane of the radiation is very similar to that in
\cite{kmss}. There are however additional physical effects of this
extended augmentation. The most important of these is the 
possibility of an
effective time dependent fine structure constant occurring because of
the P-conserving interaction 
\begin{eqnarray}
S_{int}^P ~=~{\zeta_- \over M_P}~\int d^4x~ \phi_H^{(+)} ~F^2~. 
\label{alpt} 
\end{eqnarray}
For a homogeneous $\phi_H^{(+)}=\phi_H^{(+)}(t)$,
the (\ref{alpt}) can be added on to the Maxwell action leading to an
effective $\alpha_{eff}(t)$ given by
\begin{eqnarray}
\alpha_{eff}^{-1}(t) ~=~\alpha_0^{-1} + {\zeta_- \over 
M_P}~\phi_H(t)~. \label{alph}
\end{eqnarray}
Given that $\phi_H^{(+)}(t) = \zeta_-(h/M_P)~t(z)$, where $t(z)$ is 
the 
lookback time, eq. (\ref{alph}) implies that the effective fine 
structure constant was smaller in the past. Using upper limits 
obtained recently on the time variation of the fine structure 
constant \cite{webb}, one may bound the parameter $\zeta_-h$ from 
above. This implies that the appearance of P-violating multipole 
moments in the CMB anisotropy will be correlated with the 
time-variation of the effective fine structure constant, if, as per 
our proposal here, they both originate from the same P-violating 
interaction of the KR field.

One now comes to the interesting question as to what happens when 
the P-violating interaction described above is superimposed on the 
warp factor-enhanced KR-Maxwell Chern Simons coupling in the RS 
scenario obtained in \cite{mas}. We work, following 
\cite{mas}, within a five dimensional RS scenario described by the 
metric
\begin{equation}
ds^2 = e^{-2 \sigma}~ \eta_{\alpha \beta}~ dx^{\alpha}  dx^{\beta}  
~+~ r_c^2~ d\phi^2~, \label{rsmet}
\end{equation}
where $\sigma = k r_c  \vert \phi  \vert   $. The augmentation of 
the KR field, propagating in the bulk, is given by the Maxwell 
Chern Simons three-tensor on the brane together with the extension 
proposed above
\begin{eqnarray}
\tilde H_{M N L} ~=~ \partial_{[M} B_{N L]}~ +~ {1 
\over 3M_p^{1/2}}~ \delta_M^{\mu} \delta_N^{\nu}
\delta_L^{\lambda}~A_{[\mu}~[\zeta_+ F_{\nu 
\lambda]}~+~\zeta_- {}^*F_{\nu \lambda]}]~. \label{augh}
\end{eqnarray}
This augmentation leads to the interaction
\begin{eqnarray}
S_{int} ~=~ \frac 1 {3M_p^{\frac 1 2}}~&& \int d^5 x \sqrt {-g}~
\delta_{\mu}^M\delta_{\nu}^N\delta_{\lambda}^L~\delta(\phi 
- \pi)~H_{MNL}  \nonumber \\
&& ~ A^{[\mu}~\left[ \zeta_+F^{\nu\lambda]}~+~\zeta_- 
{}^*F^{\nu \lambda]} \right] ~. \label{nint}
\end{eqnarray} 
Now, the KR potential $B_{\mu \nu}(x,\phi)$ in the bulk has the 
Kaluza-Klein decomposition
\begin{eqnarray}
B_{\mu \nu}(x,\phi) = \sum^{\infty}_{n=0} B^n_{\mu \nu}(x)
\chi^n(\phi) \frac 1{\sqrt{r_c}}~. \label{bkk}
\end{eqnarray}
The 4d effective action for $B_{\mu \nu }$ on the brane is expected 
to have the form
\begin{eqnarray}
S_H = \int d^4 x \sum^{\infty}_{n=0} \left[ \eta^{\mu \alpha}
\eta^{\nu \beta}
\eta^{\lambda \gamma}  H^n_{\mu \nu \lambda} H^n_{\alpha \beta
\gamma} + 3 m^2_n
\eta^{\mu \alpha} \eta^{\nu \beta} B^n_{\mu \nu} B^n_{\alpha 
\beta}\right]~,
\end{eqnarray}
where the modes $\chi^n$ satisfy
\begin{equation}
- \frac 1 {r^2_c}  \frac {\partial^2 \chi^n} {\partial \phi^2} =
m^2_n \chi^n e^{2 \sigma}~. \label{kksol}
\end{equation}
Following \cite{weprl}, we infer the existence of a constant zero mode 
\begin{equation}
\chi^0 = \sqrt {k r_c} e^{-k r_c \pi}~.\label{zer}
\end{equation}
Using eq. (\ref{nint}), one obtains the interaction 
\begin{eqnarray}
S_{int} ~=~ \frac {r_c} {M_p^{\frac 1 2}} &&~ \int  d^4x~ d\phi ~
e^{2\sigma}~ \delta(\phi-\pi) ~
\eta^{\mu\rho}\eta^{\nu\sigma}\eta^{\lambda\xi}~  
H_{\mu \nu \lambda}(x,\phi)~\nonumber \\
& {\mathrm X}&~A_{[\rho}~\left[\zeta_+ F_{\sigma \xi]} ~+~\zeta_- 
{}^*F_{\sigma \xi]}\right]~. \label{sint}
\end{eqnarray}
Observe that the contraction of the KR field strength with the 
Maxwell three-tensor fields has entailed the use of several inverse 
four-metric components which has resulted in not only cancelling all 
the warp factors, but also in an extra {\it inverse} warp factor 
leftover in the integrand in (\ref{sint}). Integrating over the 
bulk coordinate $\phi$  and using eq. (\ref{zer}) one obtains the 
effective interaction on the brane
\begin{equation}
S_{int} = \sqrt {\frac k M_p}r_c e^{k r_c \pi}  \int d^4 x
 H_{\mu \nu \lambda}(x) A^{[\mu}\left[ \zeta_+ F^{\nu \lambda]} + 
\zeta_- {}^*F^{\nu \lambda]} \right]~. \label{snt}
\end{equation}
Replacing the KR field strength by its dual $\phi_H$ and using the 
decomposition (\ref{dec}) into parity eigenstates, one obtains the 
braneworld version of eq. (\ref{pvio})
\begin{eqnarray}
S_{int}^{\mathrm PV} ~=~\sqrt{k r_c^2\over M_P}~\exp\{ \pi k r_c 
\}~\zeta_+ \int \phi_H^{(+)}~F \cdot ^*F~. 
\label{pvio2}
\end{eqnarray}
Just as in \cite{mas}, the optical rotation due to this interaction 
will be huge, in view of the exponential enhancement factor $\exp 
\{ \pi kr_c \}$, and from eq. (\ref{pvio2}) therefore one expects 
{\it 
large} P-violating multipole moment correlations from eq. 
(\ref{pvcor}) in the CMB anisotropy. Once again, the only way to 
control this runaway enhancement will be to fine tune $h$ to 
unnaturally small values\footnote{First numerical estimates on how 
precisely one must fine tune $h/M_P$ are in the $O(10_{15})$ range; 
this will be elaborated elsewhere.}. 

The situation is equally compelling for the time variation of the
fine structure constant; to obtain the observed relative variation
of $O(10^{-5})$ per year \cite{webb}, unnatural fine tuning of 
$\zeta_-h$ seems imperative. We have assumed that $\zeta_-=O(1)$ so 
that reconciling with observations amounts to unnatural fine tuning 
of $h$. 

To conclude, starting from an extended augmentation of the KR field
strength beyond the string theory motivated Chern Simons
augmentation, and having a different spatial parity from the latter,
and using a five dimensional RS framework as our setting, we have
demonstrated the possibility of occurrence of abnormally large 
P-violating temperature-polarization multipole moment correlations 
in the CMB anisotropy. The model also predicts abnormally large time 
variations of the fine structure constant. Both these require fine 
tuning of the KR parameter $h/M_P$ to {\it unnaturally} small values 
in order to be reconciled with (non-)observations.

In conclusion this work shows a possible difficulty one may encounter to 
establish a connection between stringy KR background and Randall-Sundrum braneworld.
However keeping in mind that a proper stringy realization of Randall-Sundrum braneworld model is yet to
be formulated, our work may play a crucial role in finding this connection.

One of us (SS) thanks M. Kamionkowski for stimulating
correspondence related to possible generalization of the work in
\cite{mas} to the case of CMB anisotropy. Another of us (PM) thanks
D. Bailin, F. Barbero, S. Bharadwaj, A.  Chatterjee, T. Souradeep,
S. Theisen and S. Umasankar for discussions, and the Theory Group at
CERN, the Theoretical Physics Group at IMAFF, the Department of
Physics at the University of Sussex and the Albert Einstein
Institute for hospitality during the various stages of completion of
this work. DM acknowledges the Council for Scientific and Industrial Research, Government of
India for financial support.

\end{document}